\documentclass[twocolumn,showpacs,jcp,superscriptaddress]{revtex4}
\usepackage{graphicx}
\usepackage[usenames]{color}
\usepackage{amssymb}
\usepackage{amsmath}
\usepackage{amsfonts}
\usepackage{mathrsfs}

\renewcommand{\epsilon}{\varepsilon}
\newcommand{\figurewidth}{0.46\textwidth}
\newcommand{\narrowfigurewidth}{0.25\textwidth}

\begin{document}
\title{Polymer translocation into laterally unbounded confined environments}

\author{Kaifu Luo}
\altaffiliation[]{Author to whom the correspondence should be addressed}
\email{kluo@ustc.edu.cn}

\affiliation{CAS Key Laboratory of Soft Matter Chemistry, Department of Polymer Science and Engineering, Hefei National Laboratory for Physical Sciences at the Microscale, University of Science and Technology of China, Hefei, Anhui Province 230026, P. R. China}
\affiliation{Physics Department, Technical University of Munich,
D-85748 Garching, Germany}


\author{Ralf Metzler}
\altaffiliation[]{Author to whom the correspondence should be addressed}
\email{metz@ph.tum.de}
\affiliation{Physics Department, Technical University of Munich,
D-85748 Garching, Germany}

\date{\today}

\begin{abstract}

Using Langevin dynamics simulations in three dimensions (3D), we
investigate the dynamics of polymer translocation into the regions between two
parallel plane walls with separation $R$ under
a driving force $F$, respectively. Compared with an unconfined environment, the
translocation dynamics is greatly changed due to the crowding effect of the
partially translocated monomers.
Translocation time $\tau$ initially decreases rapidly with increasing $R$ and then
saturates for larger $R$, and the confined environment leads
to a nonuniversal dependence of $\tau$ on $F$.

\end{abstract}


\maketitle

\section{Introduction}

The transport of biopolymers through a nanopore has attracted broad interest
because it is a challenging problem in polymer physics and is also related to
many biological processes, such as DNA and RNA translocation across nuclear
pores, protein transport through membrane channels, and virus injection.
Due to its potentially revolutionary technological
applications \cite{Kasianowicz,Meller03}, including rapid DNA sequencing, gene
therapy and controlled drug delivery, a number of recent
experimental \cite{Akeson,Meller00,Meller01,Bashir,Bayley,Branton,Storm05} and
theoretical \cite{Storm05,Sung,Park,Muthukumar99,MuthuKumar03,Lubensky,Kafri,
Ambj,Chuang,Kantor,Kantor2,Dubbeldam1,Luo1,Luo2,Huopaniemi1,Huopaniemi2,Luo3,Luo4,
Luo5,Luo7,Luo6,Luo8,Aniket,Sakaue,Slater,Liao,Lu}
studies have been devoted to this subject.

From both the basic physics as well as a technology design perspective, an important measure
is the scaling of the average translocation time $\tau$ with the polymer length $N$,
$\tau\sim N^{\alpha}$, and the value of the corresponding scaling exponent $\alpha$.
Standard equilibrium Kramers analysis
\cite{Kramers} of diffusion through an entropic barrier yields $\tau \sim N^2$
for unbiased translocation and $\tau \sim N$ for driven translocation (assuming
friction to be independent of $N$) \cite{Sung,Muthukumar99}.
However, as noted by Chuang \textit{et al}. \cite{Chuang} the quadratic scaling
behavior for unbiased translocation cannot be correct for a self-avoiding
polymer. The reason is that the translocation time would be shorter than the Rouse
equilibration time of a self-avoiding polymer, $\tau_R \sim N^{1+2\nu}$, where
the Flory exponent $\nu=0.588$ in 3D and $\nu_{2D}=0.75$ in 2D \cite{deGennes,Doi,Rubinstein}.
This observation renders the concept of equilibrium entropy and the ensuing entropic
barrier inappropriate for polymer translocation dynamics. Chuang \textit{et al}.
\cite{Chuang} performed numerical simulations with Rouse dynamics for a 2D
lattice model to study the translocation for both phantom and self-avoiding
polymers. They decoupled the translocation dynamics from the diffusion dynamics
outside the pore by imposing the restriction that the first monomer, which is
initially placed in the pore, is never allowed to escape back out of the pore.
Their results show that for large $N$, $\tau \sim N^{1+2\nu}$, which scales in
the same manner as the equilibration time but with a considerably larger prefactor.
This result was recently corroborated by extensive numerical simulations based
on the Fluctuating Bond (FB) \cite{Luo1} and Langevin Dynamics (LD) models with
the bead-spring approach \cite{Huopaniemi1,Liao}.
For driven translocation, Kantor and Kardar \cite{Kantor} have demonstrated
that the assumption of equilibrium in polymer dynamics in previous work
\cite{Sung,Muthukumar99} breaks down more easily and
provided a lower bound $\tau \sim N^{1+\nu}$ for the translocation time by
comparison to the unimpeded motion of the polymer. Using FB \cite{Luo2} and LD
\cite{Huopaniemi1,Luo3} models, a crossover from $\tau \sim N^{2\nu}$ for
relatively short polymers to $\tau \sim N^{1+\nu}$ for longer chains was found
in 2D. In 3D, we find that for faster translocation processes $\tau \sim
N^{1.37}$ \cite{Luo6,Luo7}, while it crosses over to $\tau \sim N^{1+\nu}$ for
slower translocation, corresponding to weak driving forces and/or high
friction \cite{Luo8}.

However, above physical pictures are based on translocation into an unconfined
\textit{trans} side. Very little attention is paid to the dynamics of
translocation into confined environments. It is known that the crowding due to
macromolecular aggregates and other inclusions in the cellular cytoplasm can be
as high as 50\% by volume and has considerable influence on reaction rates,
protein folding rates, and equilibria in \textit{vivo} \cite{Review}.
Similarly, polymer translocation into a confined environment is subject to a
large entropic penalty which should dramatically affect the translocation
dynamics. In addition, studies on translocation into confined geometries will
shed light on the dynamics of DNA packaging \cite{Packaging}.

We here
investigate the dynamics of polymer translocation into the region between two
parallel plane walls (3D) using Langevin dynamics simulations.
The paper is organized as follows. In section II, we briefly describe our model
and the simulation technique. In section III, we present our results. Finally,
the conclusions and discussion are presented in section IV.

\section{Model and methods} \label{chap-model}

In our numerical simulations, the polymer chains are modeled as bead-spring
chains of Lennard-Jones (LJ) particles with the Finite Extension Nonlinear
Elastic (FENE) potential. Excluded volume interaction between monomers is
modeled by a short range repulsive LJ potential: $U_{LJ} (r)=4\epsilon
[{(\frac{\sigma}{r})}^{12}-{(\frac{\sigma} {r})}^6]+\epsilon$ for $r\le
2^{1/6}\sigma$ and 0 for $r>2^{1/6}\sigma$. Here, $\sigma$ is the diameter of a
monomer, and $\epsilon$ is the depth of the potential. The connectivity between
neighboring monomers is modeled as a FENE spring with $U_{FENE}
(r)=-\frac{1}{2}kR_0^2\ln(1-r^2/R_0^2)$, where $r$ is the distance between
consecutive monomers, $k$ is the spring constant and $R_0$ is the maximally
allowed separation between connected monomers.

\begin{figure}
  \includegraphics*[width=\narrowfigurewidth]{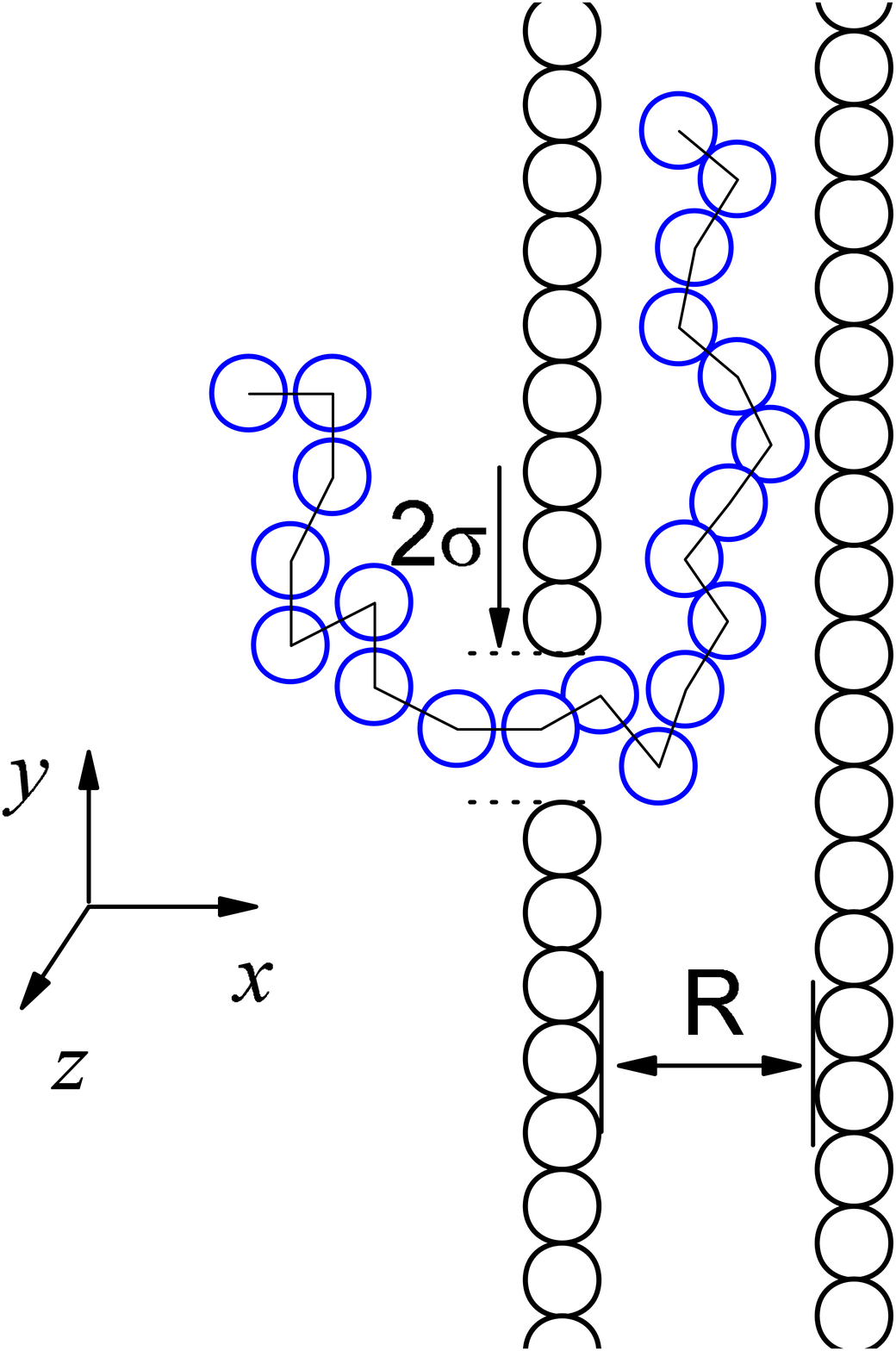}
\caption{Schematic representation of polymer translocation into a
confined environment under an external driving force $F$ in the pore. The
simulations are carried out in a planar confinement (3D), where two plates are
separated by a distance $R$. One plate has a pore of length
$L=\sigma$ and diameter $W=2\sigma$.
        }
 \label{Fig1}
\end{figure}

We consider a geometry as shown in Fig. \ref{Fig1}, where two walls with
separation $R$ are formed by stationary particles within a distance $\sigma$
from each other. One wall has a pore of diameter $2\sigma$.
Between all monomer-wall particle pairs, there exist the same short range
repulsive LJ interaction as described above.
In the Langevin dynamics simulation, each monomer is subjected to conservative,
frictional, and random forces, respectively, with \cite{Allen} $m{\bf \ddot
{r}}_i =-{\bf \nabla}({U}_{LJ}+{U}_{FENE})+{\bf F}_{\textrm{ext}} -\xi {\bf
v}_i + {\bf F}_i^R$, where $m$ is the monomer's mass, $\xi$ is the friction
coefficient, ${\bf v}_i$ is the monomer's velocity, and ${\bf F}_i^R$ is the
random force which satisfies the fluctuation-dissipation theorem.
The external force is expressed as ${\bf F}_{\textrm{ext}}=F\hat{x}$, where $F$
is the external force strength exerted exclusively on the monomers in the pore, and
$\hat{x}$ is a unit vector in the direction along the pore axis.

In the present work, we use the LJ parameters $\epsilon$ and $\sigma$ and the
monomer mass $m$ to fix the energy, length and mass scales respectively. The time
scale is then given by $t_{LJ}=(m\sigma^2/\epsilon)^{1/2}$. The dimensionless
parameters in our simulations are $R_0=2$, $k=7$, $\xi=0.7$ and $k_{B}T=1.2$
unless otherwise stated. The driving force $F$ is set between $1.0$ and $15$,
which corresponds to the range of voltages used in the
experiments \cite{Kasianowicz,Meller01}. The Langevin equation is integrated in
time by a method described by Ermak and Buckholz \cite{Ermak} in 3D.

Initially, the first monomer of the chain is placed in the entrance of the
pore, while the remaining monomers are undergoing thermal collisions described by
the Langevin thermostat to obtain an equilibrium configuration.
Typically, we average our data over 1000 independent runs.

\section{Results and discussion} \label{chap-results}


\begin{figure}
\includegraphics*[width=\figurewidth]{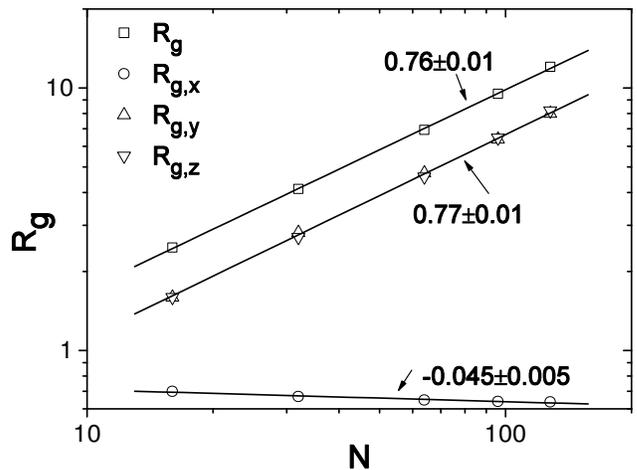}
\caption{The radius of gyration at equilibrium state as a function of the
chain length confined between two walls with separation $R=3.5$.
In the figure, $R_{g,y}=R_{g,z}=R_{\parallel}$ is the direction parallel to the wall,
while $R_{g,x}=R_{\perp}$ is the direction perpendicular to the wall.
        }
\label{Fig2}
\end{figure}

\begin{figure}
\includegraphics*[width=\figurewidth]{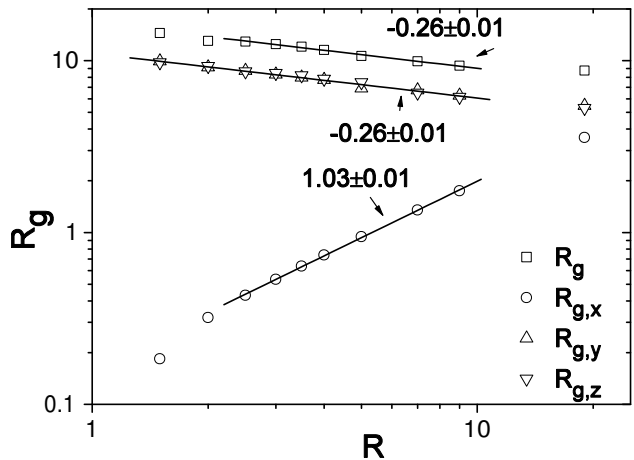}
\caption{The radius of gyration at equilibrium state under the confinementi
between two walls with
separation $R$ for $N=128$. In the figure, $R_{g,y}=R_{g,z}=R_{\parallel}$ is
in the direction parallel to the wall,
while $R_{g,x}=R_{\perp}$ is in the direction perpendicular to the wall.
        }
\label{Fig3}
\end{figure}

According to the blob picture \cite{deGennes}, a chain confined between two
parallel plates with separation $\sigma \ll R \ll R_g$ will form a 2D
self-avoiding walk consisting of $n_b$ blobs of size $R$. Each blob contains
$g=(R/\sigma)^{1/\nu}$ monomers and the number of blobs is
$n_b=N/g=N(\sigma/R)^{1/\nu}$. Thus, the blob picture predicts the
longitudinal size of the polymer to be \cite{Doi,deGennes} $R_{\parallel}
\sim {n_b}^{\nu_{2D}}R \sim N^{\nu_{2D}} \sigma (\frac {\sigma}
{R})^{\nu_{2D}/\nu-1} \sim N^{3/4} \sigma (\frac {\sigma} {R})^{0.28}$, where
the Flory exponent $\nu_{2D}=0.75$ in 2D and $\nu=0.588$ in 3D.
Fig. \ref{Fig2} shows the radius of gyration as a function of the chain length
with wall separation $R=4.5$.
We obtain $R_{g,y}=R_{g,z} \sim N^{0.77\pm0.01}$, which is in good agreement with
$R_{\parallel}\sim N^{3/4}$.
$R_{g,x}$ almost does not change with $N$, but $R_g$ scales with $N$ in the same
way as $R_{\parallel}$.
The $R$ dependence of $R_{\parallel}$ is shown in Fig. \ref{Fig3}, where
$R_{g,y}=R_{g,z} \sim R^{-0.26\pm0.01}$, in good agreement with $R_{\parallel}\sim (\frac {1} {R})^{0.28}$.
For $10.0\ge R \ge 2.5$, $R_{g,x}$ shows linear behavior
with $R$, but $R_{g} \sim R^{-0.26\pm0.01}$.

The longitudinal relaxation time $\tau_{\parallel}$ is defined as the time for
a polymer moving a distance of order of its longitudinal size, $R_{\parallel}$.
Thus, $\tau_{\parallel}$ scales as $\tau_{\parallel} \sim \frac
{R_{\parallel}^2} {\widetilde{D}} \sim N^{1+2\nu_{2D}}R^{2(1-\nu_{2D}/\nu)}
\sim N^{2.50}R^{-0.55}$, where $\sigma$ is the segment length and
$\widetilde{D}\sim 1/N$ is the diffusion constant.
The free energy cost of the confined
chain in units of $k_BT$ is simply the number of the blobs,
$\mathcal{F}=N(\sigma/R)^{1/\nu}$.

For polymer translocation into confined environments, a driving force is
necessary to overcome the entropic repulsion $f(R)$ exerted by already
translocated monomers. Due to the highly non-equilibrium property
of the translocation process, it is difficult to estimate the resisting force
$f(R)$. For polymer translocation into the region between two parallel plane
walls with separation $R$, we assume that the resisting force
$f(R)$ scales as $f(R)=CR^{\gamma}$ for slow translocation processes, with $C$ and
$\gamma$ being the prefactor and the scaling exponent, respectively.
Then, under an external driving force
$F$ in the pore the translocation time $\tau$ can be written as $\tau \sim
\frac{N^{\alpha}}{F-f(R)} \sim \frac{N^{\alpha}}{F(1-CR^{\gamma}/F)}$ with
$\alpha$ being the scaling exponent of $\tau$ with chain length $N$. Due to
$\tau_{\infty} \sim \frac{N^{\alpha}}{F}$ for the unconfined system with $R \sim
\infty$ for slow translocation processes, we have
$1-\frac {\tau_{\infty}}{\tau} \sim CR^{\gamma}/F$. Based on
this relationship, we can examine the dependence of $\tau$ on $R$.


\subsection{Translocation times as a function of the chain length}

\begin{figure}
\includegraphics*[width=\figurewidth]{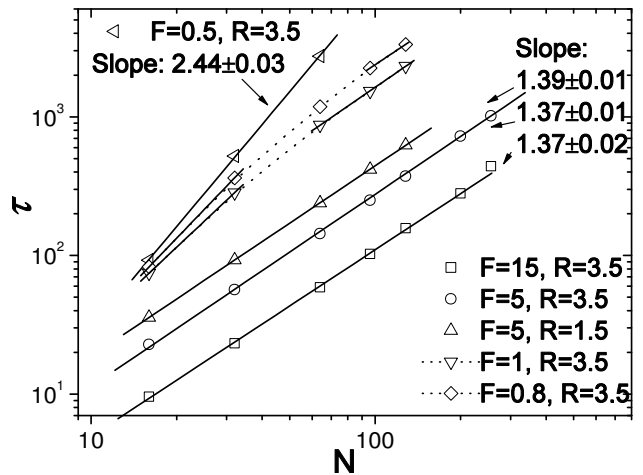}
\caption{Translocation time $\tau$ as a function of the chain length $N$ for
different $R$ and $F$ in 3D.
        }
\label{Fig4}
\end{figure}

\begin{figure}
\includegraphics*[width=\figurewidth]{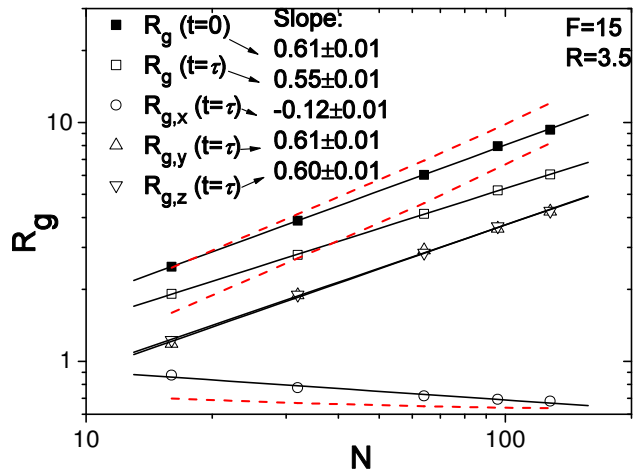}
\caption{The radius of gyration of the chain before translocation ($t=0$) and at the moment just after the translocation ($t=\tau$) for $F=15$ and $R=3.5$. Here, the $x$ direction is perpendicular to the wall, while $y$ and
$z$ are along the wall. The dashed lines are equilibrium values of  $R_g$, $R_{\parallel}$ and $R_{x}$ for chains confined between two walls, as shown in Fig. \ref{Fig2}.
        }
\label{Fig5}
\end{figure}

Fig. \ref{Fig4} shows the translocation time $\tau$ as a function of the chain
length $N$. For a strong driving force $F=5$, we find $\tau \sim N^{\alpha}$
with $\alpha=1.39\pm 0.01$ and $1.37\pm 0.01$ for $R=3.5$ and 1.5,
respectively, which are quite close to that for an unconfined system
($R=\infty$) \cite{Luo6,Luo8,Aniket,Sakaue}. For the case with very strong driving force $F=15$ and $R=3.5$,
the scaling exponent also does not change, $\alpha=1.37\pm 0.02$.
We have checked the translocation velocity with respect to the last monomer and found
$v \sim N^{\beta}$ with $\beta \approx -0.8$, which is also the same as that for an
unconfined system \cite{Luo8,Aniket}.
Under a planar confinement, during the translocation process the translocated monomers
cannot have time to diffuse away from the pore corresponding to a pronounced non-equilibrium
situation. After the translocation the
distance the last monomer has moved is always the radius of the gyration
of a chain  $R_g \sim N^{\nu}\sigma$ for its unconfined state. Thus, the
translocation time can be estimated by $\tau \sim \frac {R_g}{v}$, which is confirmed
by our numerical results.

\begin{figure}
\includegraphics*[width=\figurewidth]{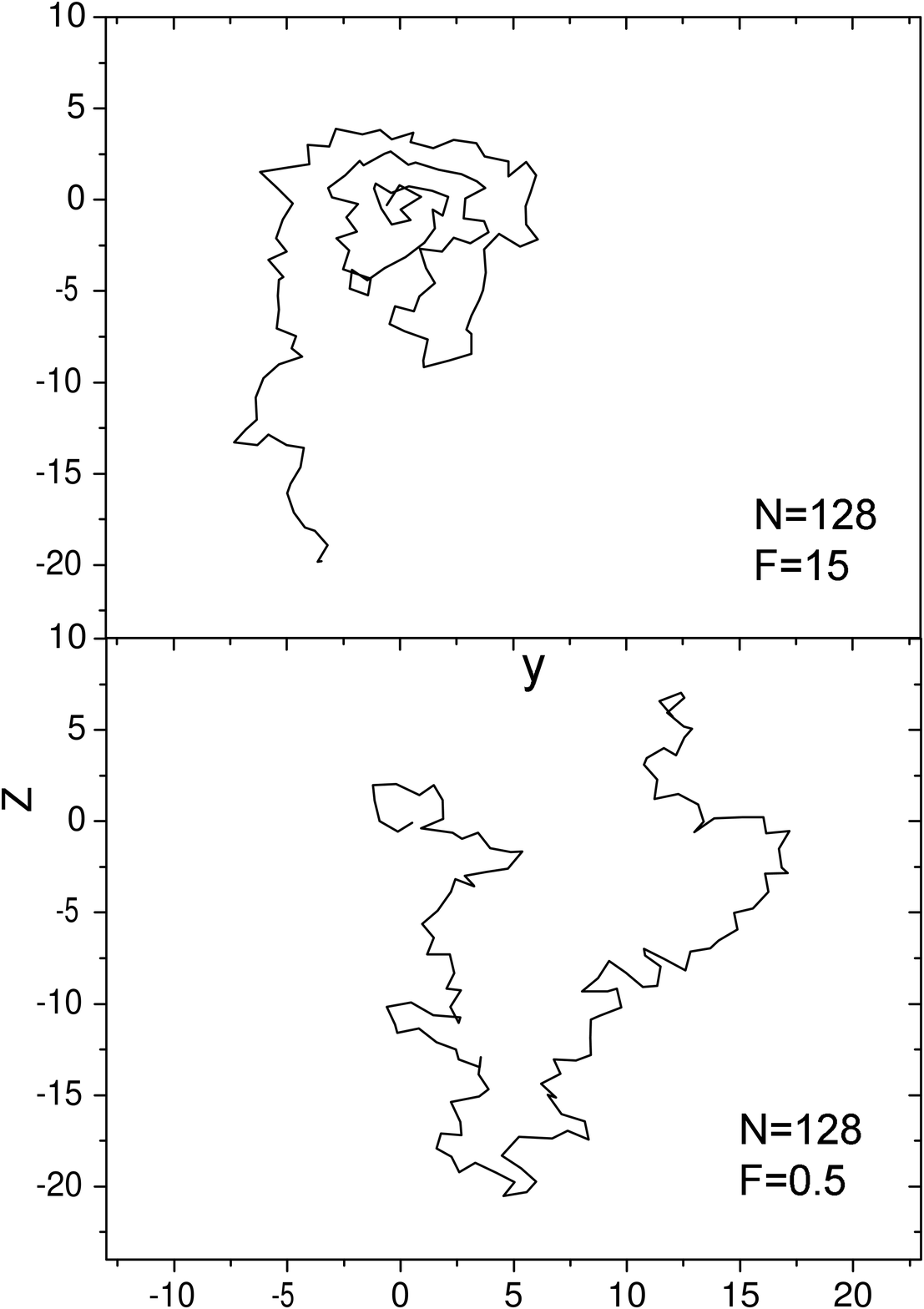}
\caption{The chain conformation projected onto the $yz$ plane for $N=128$ under strong and weak driving forces, respectively.
        }
\label{Fig6}
\end{figure}

Above scaling behavior can be understood by measuring the radius of gyration of the chain
before translocation and at the moment just after the translocation, as shown in Fig. \ref{Fig5}.
For the chain at the moment just after the translocation, $R_g\sim N^{0.55}$ and
$R_{g\parallel}\sim N^{0.60}$, significantly lower than their equilibrium values and different from
the equilibrium scaling of $N^{3/4}$ as shown in Fig. \ref{Fig2}.
These results indicate a pronouncedly non-equilibrium compression of the chain immediately after
translocation, as shown in Fig. \ref{Fig6}.

With decreasing the driving force to $F=1.0$ and $0.8$, the scaling exponent remains
the same only for longer chains.
For shorter chains $N \le 32$, the translocation is much faster due to considerably
lower confinement.

Decreasing the driving force further to $F=0.5$, we find the translocation
dynamics crosses over to another regime with $\alpha=2.44\pm 0.03$ at least for
$N \le 64$. For longer chains, the computation is very expensive due to lower
translocation probability and a longer translocation time.
This result is completely different compared with the case without the planar
confinement ($R=\infty$), where we have found $\tau \sim N^{1+\nu}$ for $F=0.5$
in our previous studies \cite{Luo8}.
Without the planar confinement, the previous prediction by Chuang \textit{et al.} \cite{Chuang}
shows that the translocation time scales in the same way as the relaxation time of the chain for unbiased translocation.
Here, our exponent $\alpha=2.44\pm 0.03$ for $F=0.5$ is very close to 2.50, which
demonstrates that $\tau$ scales in the same way as the 2D longitudinal relaxation
time for the weak driving force. The reason is that the weak driving force $F$ is almost
balanced by the resisting force due to the confinement entropy and the net driving force
is close to zero. Compared with strong driving cases, the chain conformation is not so compressed,
as shown in Fig. \ref{Fig6}.
Here, we also should point out that the exponent $\alpha=2.44$ is a little smaller than
2.50 due to the fact that during the translocation process the untranslocated monomers are
not confined. To obtain the exponent 2.5, one should initially squeeze the whole chain
into the confined spaces between two plates and then record the time during the chain
moving a distance of order of $R_{\parallel}$.

\subsection{Translocation times as a function of $R$}

\begin{figure}
\includegraphics*[width=\figurewidth]{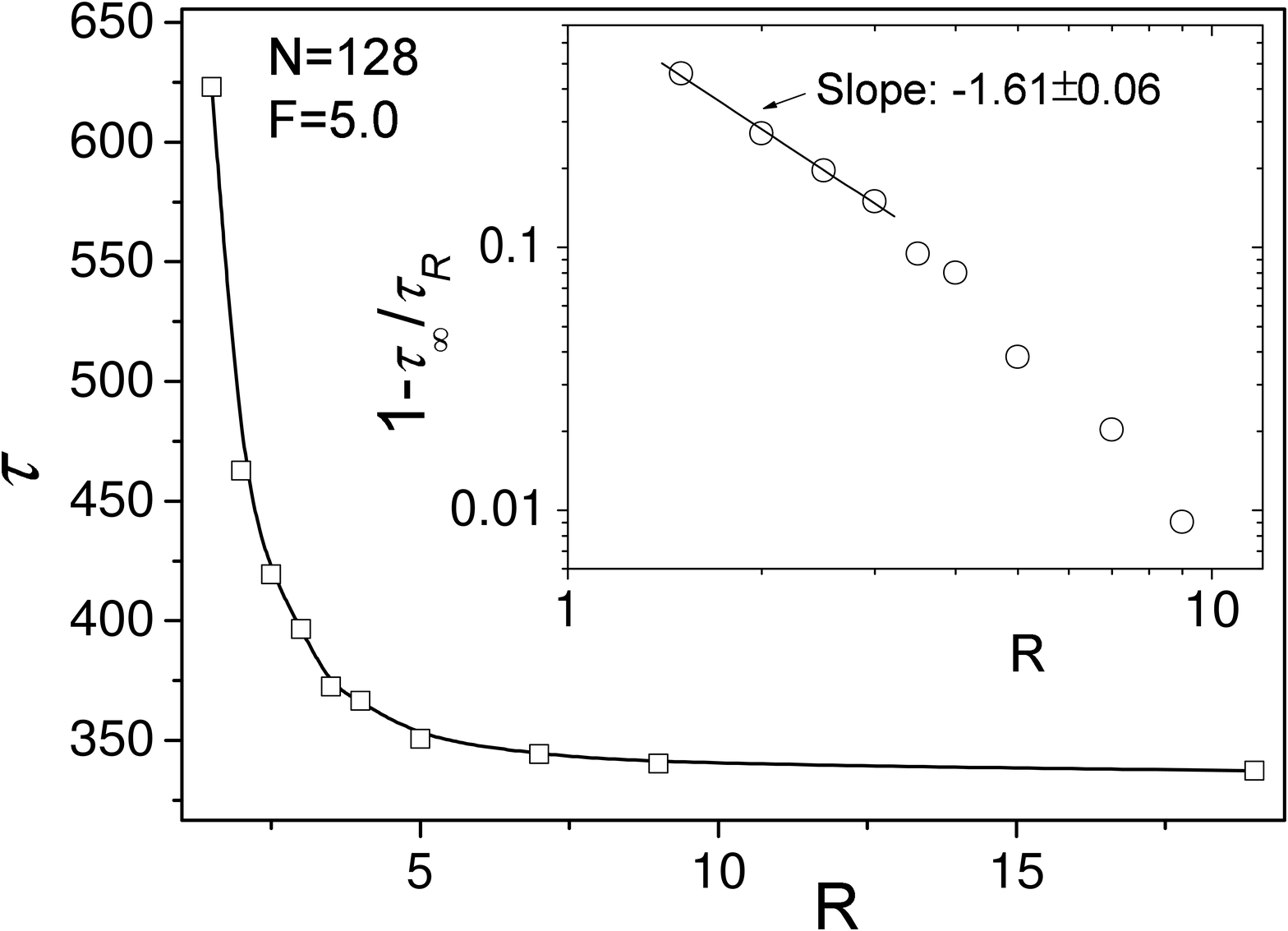}
\caption{Translocation time $\tau$ as a function of $R$ for chain length
$N=128$ under $F=5$ in 3D.
        }
\label{Fig7}
\end{figure}

\begin{figure}
\includegraphics*[width=\figurewidth]{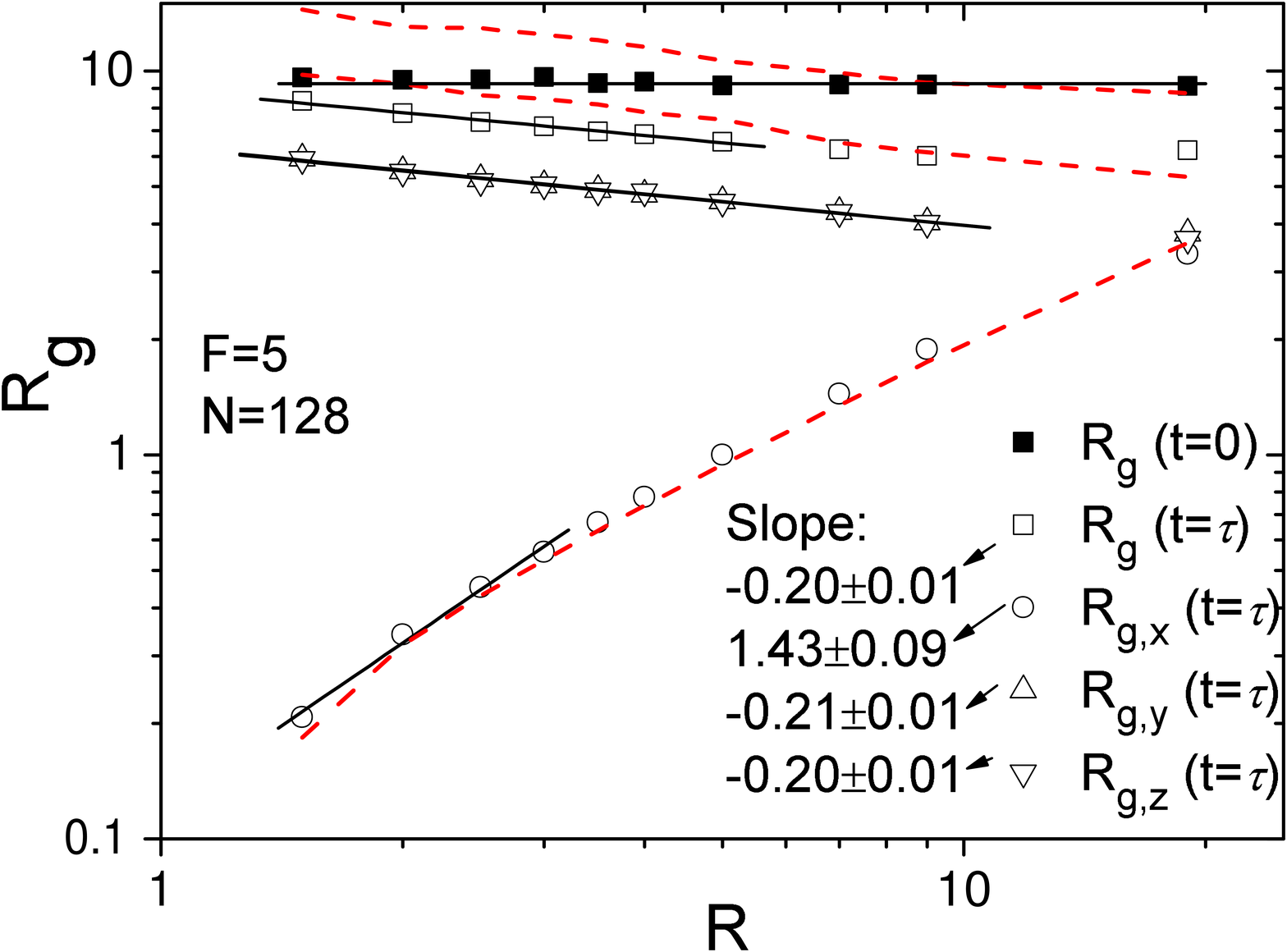}
\caption{The radius of gyration of the chain before translocation ($t=0$) and at the moment just after the translocation ($t=\tau$) for $N=128$, $F=5$ and different $R$. Here, the $x$ direction is perpendicular to the wall, and $y$ and z are along the wall. The dashed lines are equilibrium values of  $R_g$, $R_{\parallel}$ and $R_{x}$ for chains confined between two walls, as shown in Fig. \ref{Fig3}
        }
\label{Fig8}
\end{figure}

Fig. \ref{Fig7} shows the $R$ dependence of the translocation time for $F=5.0$ and
$N=128$. Initially, $\tau$ decreases rapidly with increasing $R$ and then
almost saturates for larger $R$. Quantitatively, we check $1-\frac
{\tau_{\infty}}{\tau}$ as a function of $R$, see the insert of Fig. \ref{Fig7}.
For $R \le 3$, we find $\gamma=-1.61\pm0.06$. It is close to the exponent $-1/\nu=-1.70$,
which is the theoretical value for equilibrium translocation processes with $f(R)\sim
R^{-1/\nu}$. This result also shows that for $R \le 3$, the chains really feel strong
confinement and the translocation is slow enough for the chain to be close to local equilibrium.
Fig. \ref{Fig8} shows the radius of gyration of the chain as a function of $R$, where $R_g\sim R^{-0.20}$ at the moment after the translocation is still lower than that before the translocation.
At the moment after the translocation, $R_{g,y}=R_{g,z}\sim R^{-0.20}$ for $R<9$, which is slightly lower than the equilibrium exponent $-0.28$. In addition $R_{g,x}\sim R^{1.43}$ for $R\le3$, somewhat lower than the equilibrium exponent $1/\nu=1.70$. For larger $R$, such as $R=19$, $R_{g,x}=R_{g,y}=R_{g,z}$.

\subsection{Translocation times as a function of the driving force}

\begin{figure}
\includegraphics*[width=\figurewidth]{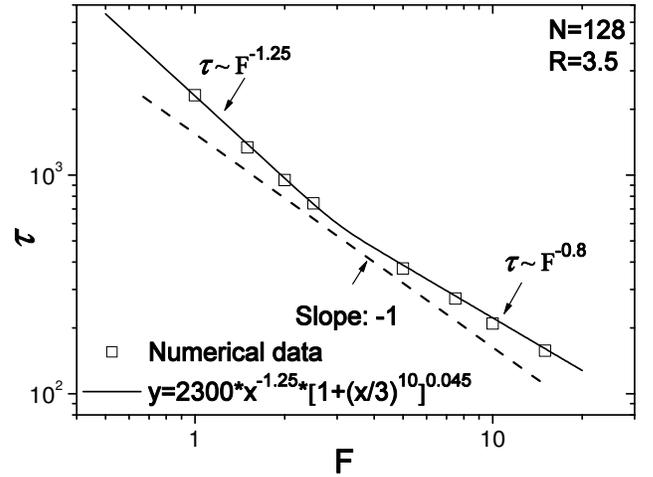}
\caption{Translocation time $\tau$ as a function of $F$ for $N=128$ and $R=3.5$
in 3D.
        }
\label{Fig9}
\end{figure}

According to the Langevin equation, when the inertia term can be neglected, the
balance of the frictional force and the driving force, $\xi v=F$, gives $\tau
\sim \xi/F$ for the case without planar confinements. The Langevin dynamics
simulation results \cite{Huopaniemi1,Luo8} confirms $\tau \sim 1/F$ only for
slow translocation with lower values of $F/\xi$. For fast translocation with
higher values of $F/\xi$, we have found $\tau \sim F^{-0.8}$ due to the highly
deformed chain conformation on the trans side, reflecting a pronounced non-equilibrium
situation \cite{Luo8}.
Fig. \ref{Fig9} shows the plot of $\tau$ as function of $F$ for translocation
into a region of a planar confinement.
Under strong driving forces, such as $F\ge 5$, compared with the case without confinement
$\tau \sim F^{-0.80\pm0.02}$ is still observed. Under a planar confinement, it is more
difficult for translocated monomers to diffuse away from the pore, which slows down the
translocation. As a result, the exponent must be lower than $-1$.
Under weak driving forces $F\le 2.5$, we find $\tau \sim F^{-1.25\pm0.03}$.
The data can be fitted with an empirical function
$\tau=2300F^{-1.25}[1+(F/3)^{10}]^{0.045}$,
yielding the details of the crossover from the weak force to the fast
strong force regime for this case.
The seemingly high exponent 10 in the empirical function is necessary to account for the
relatively fast turnover between the scaling $\tau \sim F^{-1.25}$ at weak
forces and the behavior $\tau \sim F^{-0.8}$ at strong forces, as visible in
the shown data.
For translocation
into two parallel plates with separation $R$, the effective driving force is $F(1-Cf(R)/F)$.
For weaker driving forces, the resisting force $f(R)$ becomes more and more important
because $(1-Cf(R)/F)$ is closer to zero. Therefore, with decreasing $F$ the
translocation is greatly slowed down, resulting in stronger dependence of
$\tau$ on $F$. In our previous results \cite{Luo7}, we find that the
polymer-pore interactions can also lead to this behavior.

\subsection{Waiting time distribution}

\begin{figure}
\includegraphics*[width=\figurewidth]{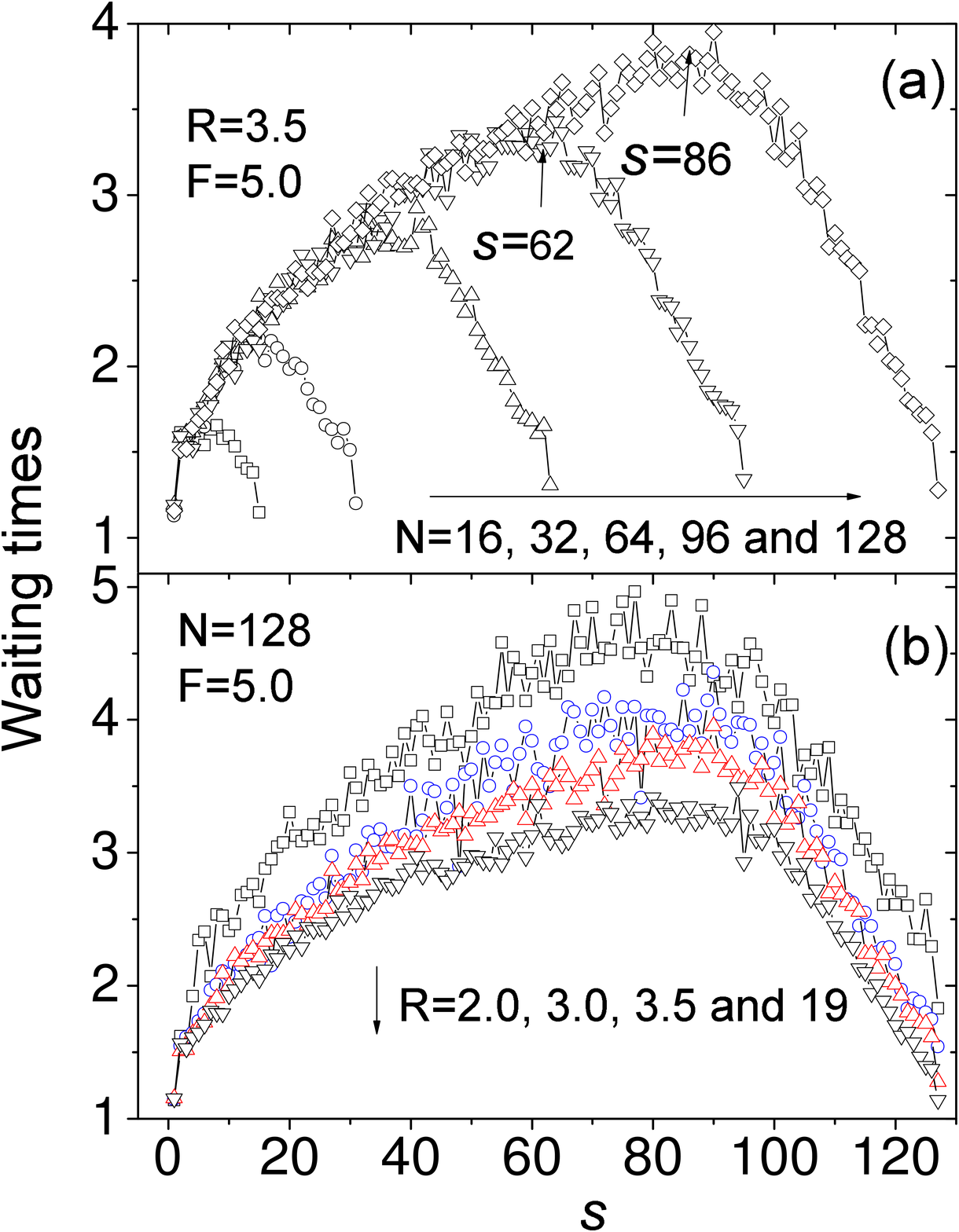}
\caption{Waiting time distribution for different (a) $N$ and (b) $R$.}
\label{Fig10}
\end{figure}

\begin{figure}
\includegraphics*[width=\figurewidth]{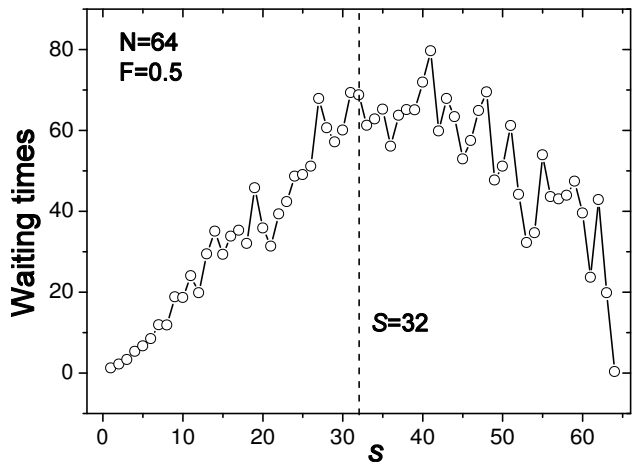}
\caption{Waiting time distribution for $F=0.5$ and $N=64$.}
\label{Fig11}
\end{figure}

The time of an individual segment passing through the pore during translocation
is an important quantity considerably affected by the translocation dynamics.
The nonequilibrium nature of translocation has a significant effect on it. We
have numerically calculated the waiting times for all monomers in a chain of
length $N$. We define the waiting time of monomer $s$ as the average time
between the events that monomer $s$ and monomer $s+1$ exit the pore.
Fig.\ref{Fig10}(a) shows the waiting time distributions for different chain lengths
for $R=3.5$ and $F=5$. One obvious feature is that before reaching their maxima
the waiting times almost follow the same pathway, which is dominated by the
strong driving force. In addition, the maxima occur at $s_{max} > N/2$, such as
$s_{max} \approx 0.65N$ and $0.67N$ for $N=96$ and 128, respectively. Fig.
\ref{Fig10}(b) shows the waiting time distributions for $N=128$ and $F=5$ for
different $R$. As expected, with decreasing $R$, the waiting times increase.
Fig.\ref{Fig11} shows the waiting time distributions for $N=64$ and $R=3.5$ under the
weak driving force $F=0.5$. After half of the polymer has been translocated, it still
takes a considerably long time to exit the pore for the remaining monomers, compared
with strong driving cases shown in Fig.\ref{Fig10}. This implies different translocation
dynamics in the two regimes.

\section{Conclusions} \label{chap-conclusions}

Using Langevin dynamics simulations, we
investigate the dynamics of polymer translocation into the region between two
parallel plane walls with separation $R$.
Compared with the chain passage into an unconfined environment, the translocation dynamics
is greatly changed. In particular, the translocation time $\tau$ initially decreases rapidly with
$R$ and then saturates for larger $R$ in 3D, and the confined
environment leads to nonuniversal dependence of $\tau$ as a function of the driving
force.
For polymer translocation into 3D confinements with  $R=3.5$, we find that
under the weak driving force $F=0.5$ the translocation time $\tau$ scales with
chain length $N$ as $\tau \sim N^{2.44}$, which is in the same manner as the
relaxation time of a chain under a planar confinement, while it crosses over to
$N$ as $\tau \sim N^{1.39}$ for fast translocation.
These behaviors are interpreted by the waiting time of individual monomer.

\begin{acknowledgments}
This work has been supported in part by the Deutsche Forschungsgemeinschaft
(DFG). K. L. acknowledges the support from University of Science and Technology of China for the candidate of Hundred Talents Program of Chinese Academy of Science.
\end{acknowledgments}

\end{document}